\documentclass[doublespacing]{elsart}
\usepackage{graphicx,graphics,amsmath,amssymb,color,lineno,subfigure}

\begin{document}
\begin{frontmatter}

\title{The atmospheric transparency measured with a LIDAR system at the Telescope Array experiment}

\author[UofYamanashi]{Takayuki Tomida\corauthref{Tomida}},
\author[UofYamanashi]{Yusuke Tsuyuguchi},
\author[UofKinki1]{Takahito Arai},
\author[UofKinki2]{Takuya Benno},
\author[UofKinki2]{Michiyuki Chikawa},
\author[UofKinki2]{Koji Doura},
\author[ICRR]{Masaki Fukushima},
\author[ICRR]{Kazunori Hiyama},
\author[UofYamanashi]{Ken Honda},
\author[ICRR]{Daisuke Ikeda},
\author[UoU]{John N.~Matthews},
\author[UofKochi]{Toru Nakamura},
\author[UofYamanashi]{Daisuke Oku},
\author[ICRR]{Hiroyuki Sagawa},
\author[TITECH]{Hisao Tokuno},
\author[ICRR]{Yuichiro Tameda},
\author[UoU]{Gordon B.~Thomson},
\author[TITECH]{Yoshiki Tsunesada},
\author[UofKanagawa]{Shigeharu Udo},
and 
\author[UofYamanashi]{Hisashi Ukai}

\corauth[Tomida]{Corresponding author. Address: Interdisciplinary Graduate 
School of Medicine and Engineering, Mechanical Systems Engineering, 
University of Yamanashi, Kofu, Yamanashi 400-8511, Japan,
Phone: +81-55-220-8418, E-MAIL:g08dm003@yamanashi.ac.jp}

\address[UofYamanashi]{Interdisciplinary Graduate School of Medicine and Engineering, Mechanical Systems Engineering, University of Ymanashi, Kofu, Yamanashi 400-8511, \quad Japan}
\address[UofKinki1]{Research Institute for Science and Technology, Kinki University, Higashiosaka, Osaka 577-8502, \quad Japan}
\address[UofKinki2]{Department of Physics, Kinki University, Higashiosaka, Osaka 577-8502, \quad Japan}
\address[ICRR]{Institute  for  Cosmic Ray Research, University of Tokyo, Kashiwa, Chiba  277-8582, \quad Japan}
\address[UoU]{Institute for High Energy Astrophysics and Department of Physics, University of Utah, Salt Lake City, UT 84112-0830, \quad USA}
\address[UofKochi]{Kochi University, Faculty of Science, Kochi 780-8520, \quad Japan}
\address[TITECH]{Graduate School of Science and Engineering, Tokyo Institute of Technology, Meguro, Tokyo 152-8551, \quad Japan}
\address[UofKanagawa]{Kanagawa University, Yokohama, Kanagawa 221-8686, \quad Japan}

\begin{abstract}
An atmospheric transparency was measured using a LIDAR with a pulsed UV laser (355nm) at the observation site of Telescope Array in Utah, USA.
The measurement at night for two years in $2007\sim 2009$ revealed that the extinction coefficient by aerosol at the ground level is $0.033^{+0.016}_{-0.012} \rm km^{-1}$ and the vertical aerosol optical depth at 5km above the ground is $0.035^{+0.019}_{-0.013}$.
A model of the altitudinal aerosol distribution was built based on these measurements for the analysis of atmospheric attenuation of the fluorescence light generated by ultra high energy cosmic rays.
\end{abstract}

\end{frontmatter}

\section{Introduction}
\label{sec:Intro}
\begin{figure}[b]
\begin{center}
\includegraphics[width=.9\columnwidth]{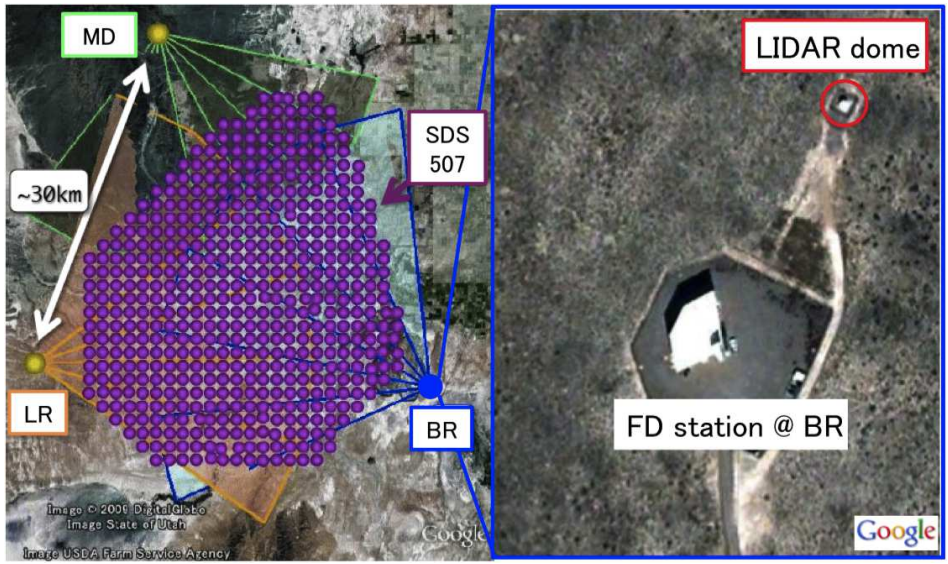}
\caption{
    Location of the LIDAR system.
    A left figure is a map of the TA experiment. Little black points are SDs, and three FD stations are shown by BR, LR, and MD.
    A right picture shows the positions of BR-station and LIDAR dome.
}
\label{lidar_location}
\end{center}
\end{figure}

Ultra High Energy Cosmic Rays (UHECRs) interact with the cosmic microwave background and produce a cutoff structure in their energy spectrum (GZK cutoff) at the energy around $10^{19.8}$ eV as predicted by Greisen, Zatsepin and Kuzmin \cite{GZK}\cite{GZK2}.
The HiRes group reported that there is a GZK cut-off in their observed energy spectrum at the energy it was predicted \cite{Hires}\cite{Hires2}.
A similar suppression in the energy spectrum is reported by the Auger experiment in the Southern Hemisphere \cite{Auger}.
However, the AGASA experiment observed a spectrum which continued unabatedly with 11 events beyond the GZK cutoff \cite{AGASA}\cite{AGASA2}.
To understand this discrepancy, the Telescope Array (TA) experiment was constructed in the desert area near the town of Delta in Utah, the USA.
It is a hybrid detector consisting of Surface Detector (SD) array and three Fluorescence Detector (FD) stations.
It measures the energy spectrum, anisotropy and composition of UHECRs to identify their origin.
The TA has three FD stations called ``Black Rock Mesa''(BR), ``Long Ridge''(LR) and ``Middle Drum''(MD), which have been installed surrounding the SD array.
The BR site (1404 m a.s.l.) is located at the southeast corner of the SD array while the LR site (1554 m a.s.l.) is to the southwest.
The BR and LR stations each have 12 telescopes.
The MD site (1610 m a.s.l.) is located at the northwest, and has 14 telescopes.
The FD stations overlook an array of 507 SDs.

The UV fluorescence light generated by an air shower is scattered and lost along the path of transmission to the telescope.
The main scattering processes are Rayleigh scattering by molecules and scattering by aerosols in the atmosphere.
The Rayleigh scattering process is well understood and the attenuation length can be calculated using the Rayleigh scattering cross-section and the molecular densities of the atmosphere \cite{Bucho_1995}\cite{Ubachs_2000}\cite{Ubachs_2005}.
In order to calculate the molecular densities of the atmosphere, radiosonde data from Elko(Nevada) are used to obtain temperature, pressure and humidity around the TA observatory as a function of height. 
Sizes, shapes and spatial distribution of aerosols around the site are not known, and are variable with time.
Therefore, on-site monitoring of aerosols is essential in a fluorescence experiment.

In the TA, we employ a variety of measurements for atmospheric monitoring, using two laser systems and a cloud camera.
The first laser system is the LIDAR (LIght Detection And Ranging, we call LIDAR) installed near the BR station, which injects a pulsed laser light in the atmosphere and observes the back-scattered light at the same location.
LIDAR is widely used in ground based aerosol measurement.
The LIDAR system is operated before the beginning and after the end of an FD observation, twice a night.
The second laser system is located at the geographical center of the three FD stations.
It fires a vertical laser beam of $355 \rm nm$ and the scattered light by the atmosphere is observed by each fluorescence detector station.
This system is called CLF (Central Laser Facility) \cite{ICRC09_tomida}\cite{ICRC07_Udo}.
We are shooting 300 shots of laser at $10 \rm Hz$ from the CLF every 30 minutes during FD observation.
In addition, we installed an infrared CCD camera for cloud monitoring near the LIDAR system, and take pictures of the night sky every hour during FD observation \cite{ICRC09_IR}.
Furthermore, the weather monitors are installed in each of the FD stations and at the CLF to obtain the atmospheric status at the ground level.

In this paper, we report on the LIDAR system (section $\ref{sec:LIDAR_system}$), the data set and the analysis method (section $\ref{sec:Analysis_method}$), the determination of the Rayleigh scattering by the radiosonde data (section $\ref{sec:molecules}$), the result of aerosol scattering by the LIDAR data (section $\ref{sec:aerosols}$), and a model of atmospheric transparency at the TA site (section $\ref{sec:Model}$).

\section{LIDAR system}
\label{sec:LIDAR_system}
\begin{figure}[t]
\begin{center}
\subfigure[LIDAR telescope and laser]{
    \includegraphics[width=.4\columnwidth]{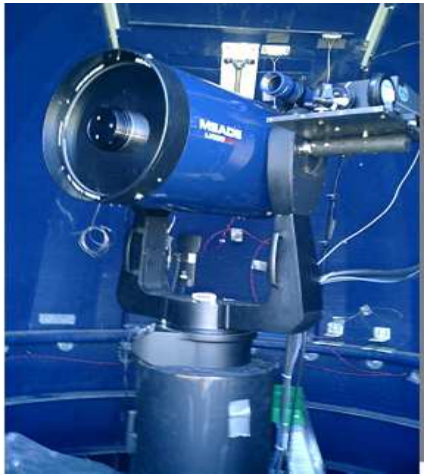}
}~
\subfigure[Block diagram of the LIDAR system]{
    \includegraphics[width=.5\columnwidth]{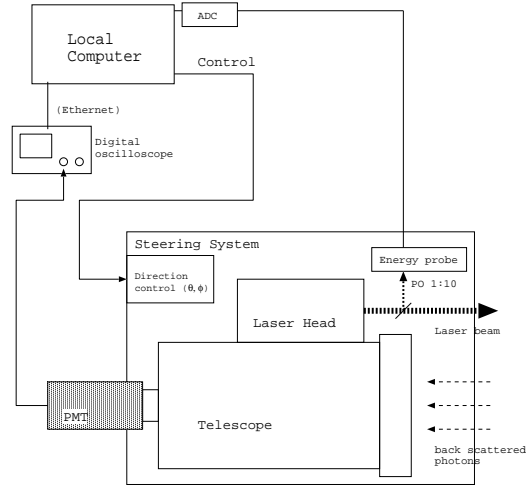}
}
\caption{
    The LIDAR system.
    A left picture is LIDAR's optical system (telescope and laser, etc.).
    A right picture is a connection block diagram of device of LIDAR.
}
\label{ext_picture}
\end{center}
\end{figure}

A photograph of the LIDAR system is shown Fig.$\ref{ext_picture}(a)$.
The LIDAR is composed of five basic system blocks. 
Those are (1) a laser, (2) a steerable telescope, (3) a photo-multiplier tube, (4) a digital oscilloscope, and (5) a control system as shown in Fig.$\ref{ext_picture}$(b).
We use an air cooled Nd:YAG laser  (Orion model by New Wave Research) with a third harmonic oscillation module ($355\rm nm$).
The laser fires a pulse with the width of $4\sim 6 \rm ns$ at $1 \rm Hz$. The maximum energy of the laser pulse is $4 \rm mJ$. 
A MEADE LX200GPS-30 telescope with a diameter of 305 mm and a focal length of 3048 mm is used with a photo-multiplier tube (HAMAMATSU R3479) mounted at the focus of the telescope.
A linear range of the PMT was checked by simultaneously firing a series of UV-LED pulses at the telescope overlaying with the laser shot, and confirming that the overlaid signal has the same charge as the LED-only signal without the laser shot.
The laser is attached to the telescope mount, therefore the direction control of the laser-telescope-PMT system can be simply accomplished by commands to the telescope mount.
The back-scattered photons from the laser are collected by the telescope, detected by the PMT, and the signals are digitized with a digital oscilloscope (Tektronix 3034B) \cite{ICRC09_tomida}\cite{ICRC07_Chikawa}.
In addition, a portion of each laser shot is picked off for measurement by an energy monitor.

By measuring the time structure of the back-scattered photons, we can determine the atmospheric conditions including aerosol distributions along the path of the laser beam.
The operation of the LIDAR is composed of horizontal shots in the north and vertical shots.
The vertical operation was made in high ($\sim 4\rm mJ$) and low ($\sim 1\rm mJ$) energies in order to measure the extinction coefficient $\alpha$ over a large range.
The horizontal operation uses only the high energy. 
In each vertical and horizontal operation, 500 laser pulses are shot and recorded.
A total of 1500 shots composed one LIDAR operation.
The pedestal level is measured in between each shot and is subtracted from the laser shot data in order to account for the background.
In the following analysis, we use an average of the PMT signal profiles in order to reduce shot-to-shot fluctuations.

\section{Data set and analysis method}
\label{sec:Analysis_method}
\begin{figure}[tb]
\begin{tabular}{cc}
\begin{minipage}{0.5\textwidth}
\begin{center}
\includegraphics[scale=0.5]{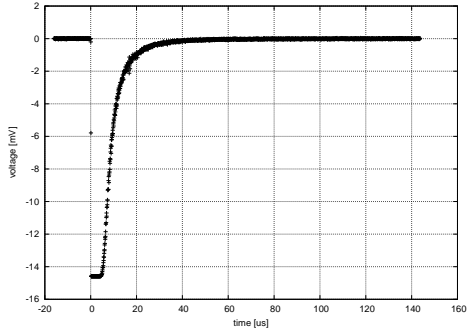}
\caption{Example of $W(t)$ for horizontal shots}
\label{oscillo}
\end{center}
\end{minipage}
\begin{minipage}{0.5\textwidth}
\begin{center}
\includegraphics[scale=0.5]{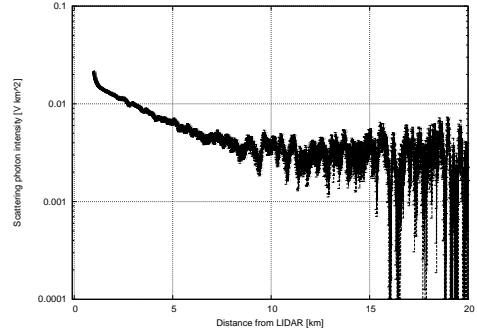}
\caption{Example of $F(x)$ as a function of the distance $x$.}
\label{intensity}
\end{center}
\end{minipage}
\end{tabular}
\end{figure}

A waveform $W(t)$ is obtained by averaging the PMT signal for 500 laser shots.
A range corrected power return $F(x)$ of the LIDAR is defined by the waveform $W(t)$ as a function of the distance $x$ from the LIDAR.
\begin{eqnarray}
F(x)=W(t)x^{2},\ \ 
x=t\frac{c}{2},
\label{eq:convert}
\end{eqnarray}
where $c$ is the speed of light and $t$ is the time from the laser shot.
The solid angle correction is taken into account in the definition of $F(x)$.
An example of $W(t)$ and $F(x)$ is shown in Fig.$\ref{oscillo}$ and Fig.$\ref{intensity}$.

The extinction coefficient $\alpha(x)$ at a distance $x$ is related with $F(x)$ by the LIDAR equation;
\begin{eqnarray}
\frac{1}{F(x)}
\frac{{\rm d}F(x)}{{\rm d}x} = \frac{1}{\beta(x)}
\frac{{\rm d}\beta(x)}{{\rm d}x} - 2 \alpha(x)
\label{eq:LIDAR_equation}
\end{eqnarray}
where $\beta$ is the back-scattering coefficient.
The factor 2 of $\alpha$ indicates a round-trip photon propagation.

If the atmosphere is spatially uniform along $x$, such as the case for the horizontal shots, $d\beta(x)/dx = 0$ and $F(x)$ becomes exponential.
We can exactly solve the equation and obtain the extinction coefficient $\alpha(x)$ (the slope method \cite{slope1}\cite{slope2}).

In the case of vertical laser shots, where the atmospheric condition ($\alpha$ and $\beta$) changes with $x$, we need an additional assumption on the relation between $\alpha$ and $\beta$ and also a boundary condition to solve the equation.
In Klett's method \cite{klett_1981}\cite{klett_1985}, a boundary condition is assumed that $\alpha$ at enough high altitude is given by that of Rayleigh scattering because the aerosol scattering can be neglected at high altitude.
The back-scattering coefficient $\beta$ is assumed to be $\beta \propto \alpha^{\kappa}$.
The value of $\kappa$ is 1.00 for the molecular atmosphere but is known to vary significantly ($0.67<\kappa<1.30$) for the atmosphere heavily loaded by the aerosol or with rain and snow \cite{klett_1985}\cite{kappa1}\cite{kappa2}\cite{kappa3}.
However, for a typical desert atmosphere of the TA site, we found that the aerosol contribution increases towards the lower altitude and that it modifies the value of $\kappa$ from 1.00 at the high altitude up to 1.14 near the ground.
In solving Eq. ($\ref{eq:LIDAR_equation}$) by Klett's method, we assume that $\beta$ is in proportion to $\alpha$ for the whole interval of $x$.
We estimate a systematic error caused by this assumption by the iterative method (see section $\ref{sec:aerosols}$).

The extinction coefficient at the distance $x$ is thus calculated from the acquired waveform $W(t)$ by adopting the slope method for horizontal shots or the Klett's method for the vertical shots.
The obtained extinction coefficient, $\alpha_{\rm obs}$, contains effects of both attenuation processes caused by Rayleigh and aerosol scattering.
The extinction coefficient by the aerosol scattering $\alpha_{\rm AS}$ is obtained using a relation $\alpha_{\rm obs}(x)=\alpha_{\rm Ray}(x)+\alpha_{\rm AS}(x)$ where $\alpha_{\rm Ray}$ is the extinction coefficient by Rayleigh scattering.

The LIDAR operation has been performed 515 times from September 2007 to October 2009.
Out of 515 operations, we passed 325 operations to the detailed analysis after removing the operations performed under apparent bad weather or bad atmospheric conditions, with irregular forms of $F(x)$ caused by the existence of clouds and other scatters, and of hardware troubles.
Two selection criteria on the data quality, the first for the horizontal shots and the second for the vertical shots, are then applied on 325 samples.
Firstly, we discarded the data with $\alpha_{\rm obs}$ on the ground level larger than $0.1 \rm km^{-1}$ because the atmosphere is too hazy for meaningful measurement.
We are left with 250 operations after this cut. Secondly, we accepted only measurements for which the lowest viable data is below $1.0 \rm km$ and the highest viable data is above $6.0 \rm km$.
A total of 137 operations have passed the second cut.
These two conditions indicate a transparent atmosphere and the altitudinal distribution of $\alpha_{\rm obs}$ can be derived from the observed LIDAR data with high accuracy.
In section $\ref{sec:aerosols}$, we use these 137 LIDAR measurements for characterizing the aerosol scattering.

\section{Scattering by atmospheric molecules}
\label{sec:molecules}
The molecules (nitrogen and oxygen) are the main sources of photon scattering in the atmosphere.
The scattering processes of photons of wavelengths in the range between $300$ and $400$ nm \cite{300_400nm_fluorescence}, which is the sensitivity band used in air fluorescence experiments, are well described by atmospheric molecules based on the Rayleigh scattering theory.
We can calculate the attenuation of the photon intensity, or the extinction coefficient, from the Rayleigh scattering cross-section and the molecular density of the atmosphere.
The former can be calculated with an accuracy of ~1\% \cite{Ubachs_2000}  and the latter is obtained by the measured distribution of atmospheric temperature and pressure employing the equation of state \cite{Bucholtz}.

We investigated radiosonde data of several stations around the TA site to determine the atmospheric conditions.
Worldwide balloon-borne radiosonde measurements are carried out at more than 900 launch stations twice a day at 0:00 and 12:00 UTC, and various atmospheric parameters including temperature and pressure are recorded along the altitude up to 30 km above the sea level.
We examined the data from 5 radiosonde stations within 500 km range of the TA site, and found that the atmospheric parameters obtained from the Elko(Nevada) are most suitable for TA data analysis.
Elko is located about $390 \rm km$ northwest of the TA site.
It is at a similar altitude to the TA site and has similar climate, temperature, and precipitation.

\begin{figure}[tb]
\begin{center}
\includegraphics[scale=0.5]{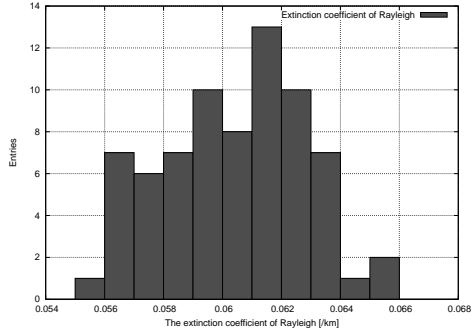}
\caption{Distribution of monthly averages of $\alpha_{\rm Ray}(0 \rm km)$ at $\lambda = 355 \rm nm$.}
\label{Rayleigh}
\end{center}
\end{figure}

In calculation of the extinction coefficients of Rayleigh scattering, we use monthly averages of the atmospheric parameters at 00:00 and 12:00.
The distribution of monthly averages of $\alpha_{\rm Ray}$ from January 2007 to December 2009 is shown in Fig.$\ref{Rayleigh}$.
The average of $\alpha_{\rm Ray}$ at the ground level and its standard deviation are $0.060\pm{0.002} \rm km^{-1}$, which translates in an average attenuation length of Rayleigh scattering of $16.7 \rm km$ or $1747 \rm g/cm^{2}$. 
The systematic error of $\alpha_{\rm Ray}$ is approximately $1\%$ coming from the inaccuracy of the Rayleigh scattering cross-section. 
At the TA site ($1400 \rm m$ a.s.l.), an average temperature of $10.2\ {}^\circ\mathrm{C}$ and atmospheric pressure of $863.3 \rm hPa$ at the ground level is measured.

\section{Scattering by the aerosol component}
\label{sec:aerosols}
The aerosol component consists of large molecules and particles of sizes up to $\sim 10^{-3}$ cm.
It influences the light scattering mostly at the low altitude.
The light scattering by a sphere of a given size is described by Mie scattering theory, however, it is difficult to apply the theory for light scattering in the lower atmosphere because we do not know shape and the spatial distribution of the aerosols a priori.

The method used in this analysis is as follows.
First, a total extinction coefficient $\alpha_{\rm obs}$ that is the sum of Rayleigh scattering and aerosol scattering is determined using the LIDAR data, in a unit of ${\rm km}^{-1}$, as a function of the height.
Next, we calculate the extinction coefficient of Rayleigh scattering $\alpha_{\rm Ray}$ using the radiosonde data.
Finally, the extinction coefficient of aerosol scattering $\alpha_{\rm AS}$ is calculated by assuming $\alpha_{\rm obs}(x) = \alpha_{\rm Ray}(x) + \alpha_{\rm AS}(x)$.

\subsection{Atmospheric transparency in the vicinity of the ground surface}
\begin{figure}[tb]
\begin{tabular}{cc}
\begin{minipage}{0.5\textwidth}
\begin{center}
\includegraphics[scale=0.5]{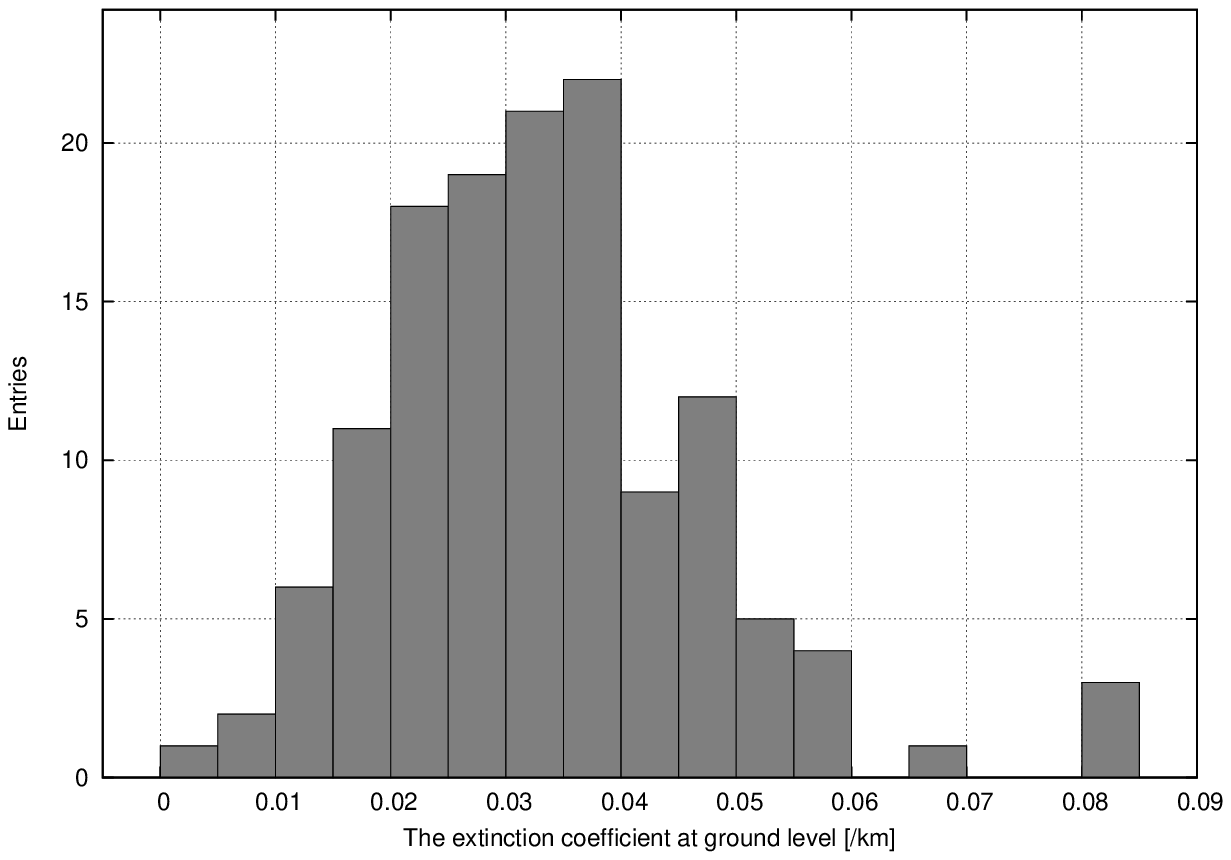}
\caption{Distribution of the $\alpha_{\rm AS}(0 \rm km)$.}
\label{slope_hist}
\end{center}
\end{minipage}
\begin{minipage}{0.5\textwidth}
\begin{center}
\includegraphics[scale=0.5]{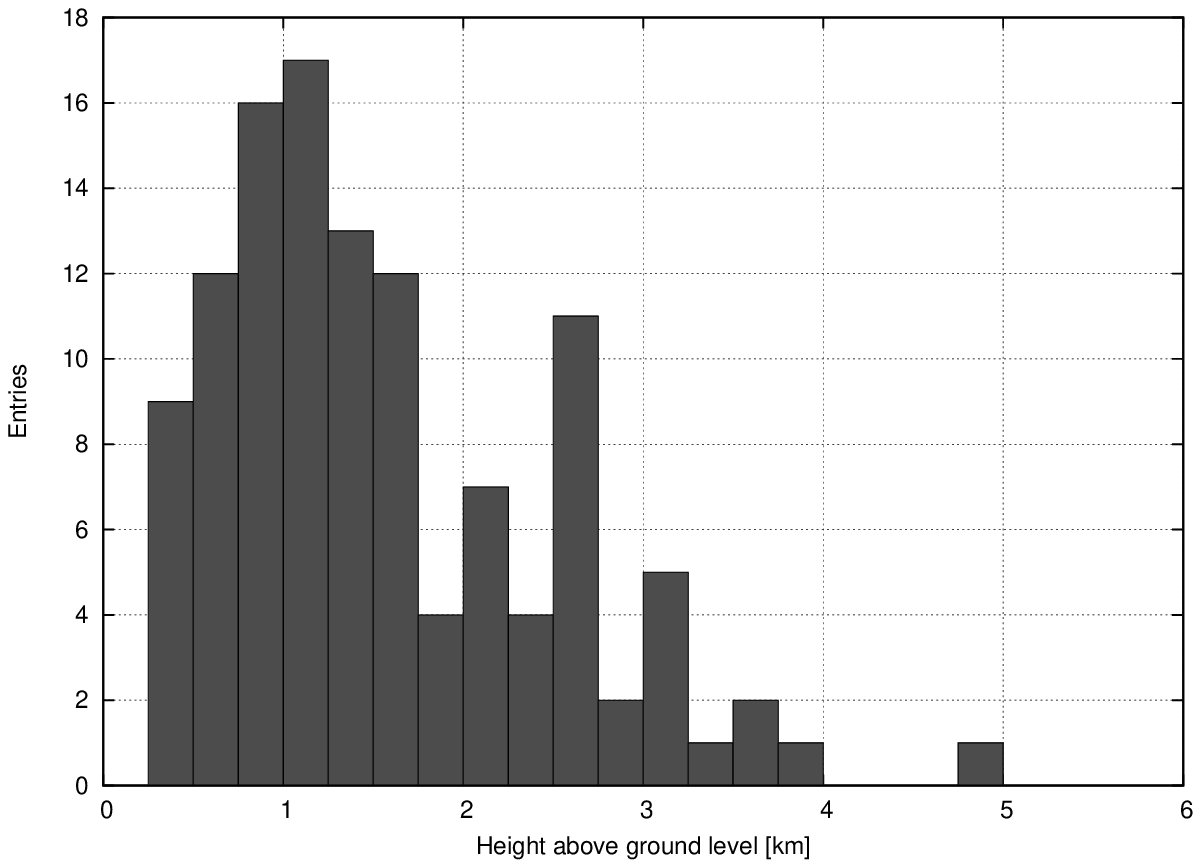}
\caption{Distribution of the maximum height below which the aerosol contribution is measurable.}
\label{Max_height}
\end{center}
\end{minipage}
\end{tabular}
\end{figure}

The aerosols are more prevalent at the ground surface and decrease as a function of the height.
They are influenced easily by the wind, ground dryness and other local factors.
Figure $\ref{slope_hist}$ shows the distribution of $\alpha_{\rm AS}$ at the ground level obtained from horizontal measurements by the slope method.
The median of $\alpha_{\rm AS}(0 \rm km)$ and its range (1$\sigma$) of distribution over 137 LIDAR operations are $0.033^{+0.016}_{-0.012} \rm km^{-1}$.
The attenuation length of aerosol scattering at the ground level was determined to be $30^{+16}_{-10} \rm km$.
The systematic error of $\alpha_{\rm AS}$ by the linearity calibration of the PMT gain is $^{+0}_{-7}\%$.

\subsection{Atmospheric transparency of each altitude}
\begin{figure}[tb]
\begin{center}
\subfigure[Little aerosol scattering]{
    \includegraphics[width=.5\columnwidth]{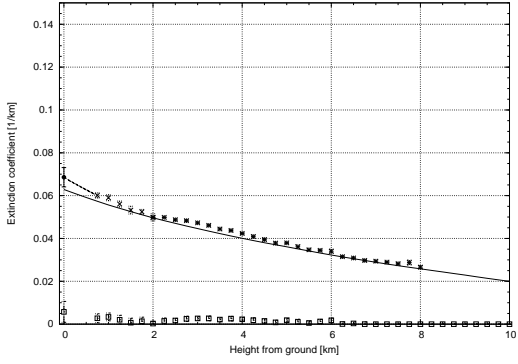}
}~
\subfigure[Aerosol distributed only at low altitude]{
    \includegraphics[width=.5\columnwidth]{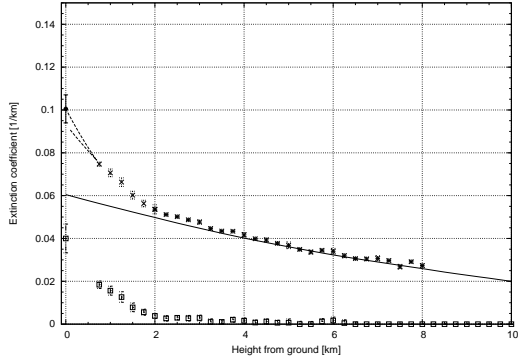}
}
\\
        \subfigure[Aerosol distributed up to high altiutde]{
            \includegraphics[width=.5\columnwidth]{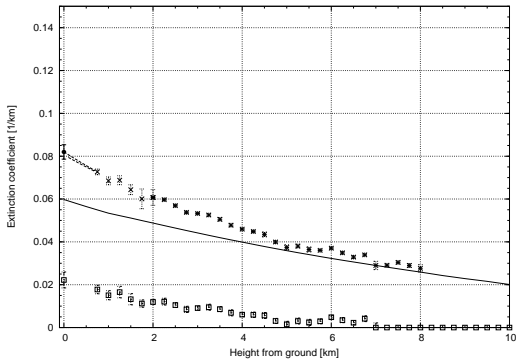}
        }~
\subfigure[Aerosol distributed at both altitudes]{
    \includegraphics[width=.5\columnwidth]{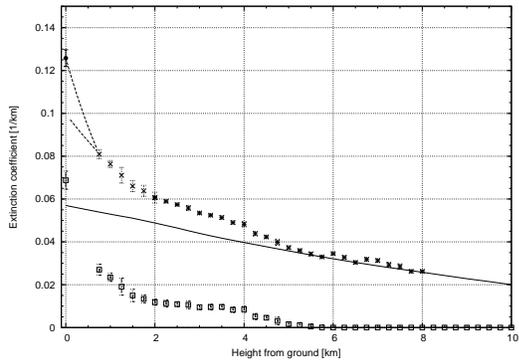}
}
\caption{
    Extinction coefficients as a function of the height from the ground level.
}
\label{ext_sample}
\end{center}
\end{figure}

The altitudinal distribution of $\alpha_{\rm obs}$ per $250 \rm m$ is calculated by Klett's method from the data of a vertical shooting.
Examples of extinction coefficient obtained from the measurement for the different distributions of aerosol scattering are shown in Fig.$\ref{ext_sample}$(a)$\sim$(d).
The $\alpha_{\rm obs}$ obtained at the ground level by the horizontal shooting and the distribution of $\alpha_{\rm obs}$ measured by the vertical shooting are shown in the closed circle and the crosses in Fig.$\ref{ext_sample}$ respectively.
For vertical shots (crosses), we used data only above $750 \rm m$ for which the linear response of the PMT was confirmed by the UV-LED calibration.
The data points above 8km are removed because the collected photon signal is too small.
The solid line of Fig.$\ref{ext_sample}$ represents $\alpha_{\rm Ray}$ calculated for pure Rayleigh scattering (see section $\ref{sec:molecules}$).
Finally the open square boxes in Fig.$\ref{ext_sample}$ represent the altitudinal distribution of $\alpha_{\rm AS}(x)$.

The systematic error of $\alpha_{\rm AS}(x)$ is $^{+0}_{-2}\%$ from the linearity calibration of the PMT, $^{+5}_{-0}\%$ from the assumption of $\kappa=1$ for Klett's method.
%The latter is estimated by iteration; at first, the aerosol contribution was obtained by the Klett's integration with the assumption of $\kappa=1$, then the effective $\kappa$ for the mixture of the molecular and aerosol scatterings was calculated ($\kappa = 1.00\sim1.06$ above $750 \rm m$ from the ground level.) and the second Klett's integration was performed.
The latter is estimated by iteration; at first, the aerosol contribution was obtained by the Klett's integration with the assumption $\beta \propto \alpha^{\kappa=1}$, then the relation between $\alpha$ and $\beta$ was recalculated at each altitude assuming the aerosol phase function obtained by the HiRes experiment \cite{HiRes2006} and the second Klett's integration was performed.
We found the second integration gives a correction of $+5\%$ or less at any altitude of the typical atmosphere.

In order to understand the aerosol contribution with respect to the height, we define a height below which the $\alpha_{\rm AS}$ is greater than $0.01 {\rm km^{-1}}$, $i.e.$ aerosol scattering is no longer negligible.
The distribution of such height is shown in Fig.$\ref{Max_height}$.
Although the distribution reaches as high as $5 \rm km$, it is less than 2 km in most cases (71\%).
We found, for $2\%$ of the cases, the contribution from the aerosol scattering was completely negligible.

\subsection{Vertical Aerosols Optical Depth}
\begin{figure}[tb]
\begin{tabular}{cc}
\begin{minipage}{0.5\textwidth}
\begin{center}
\includegraphics[scale=0.5]{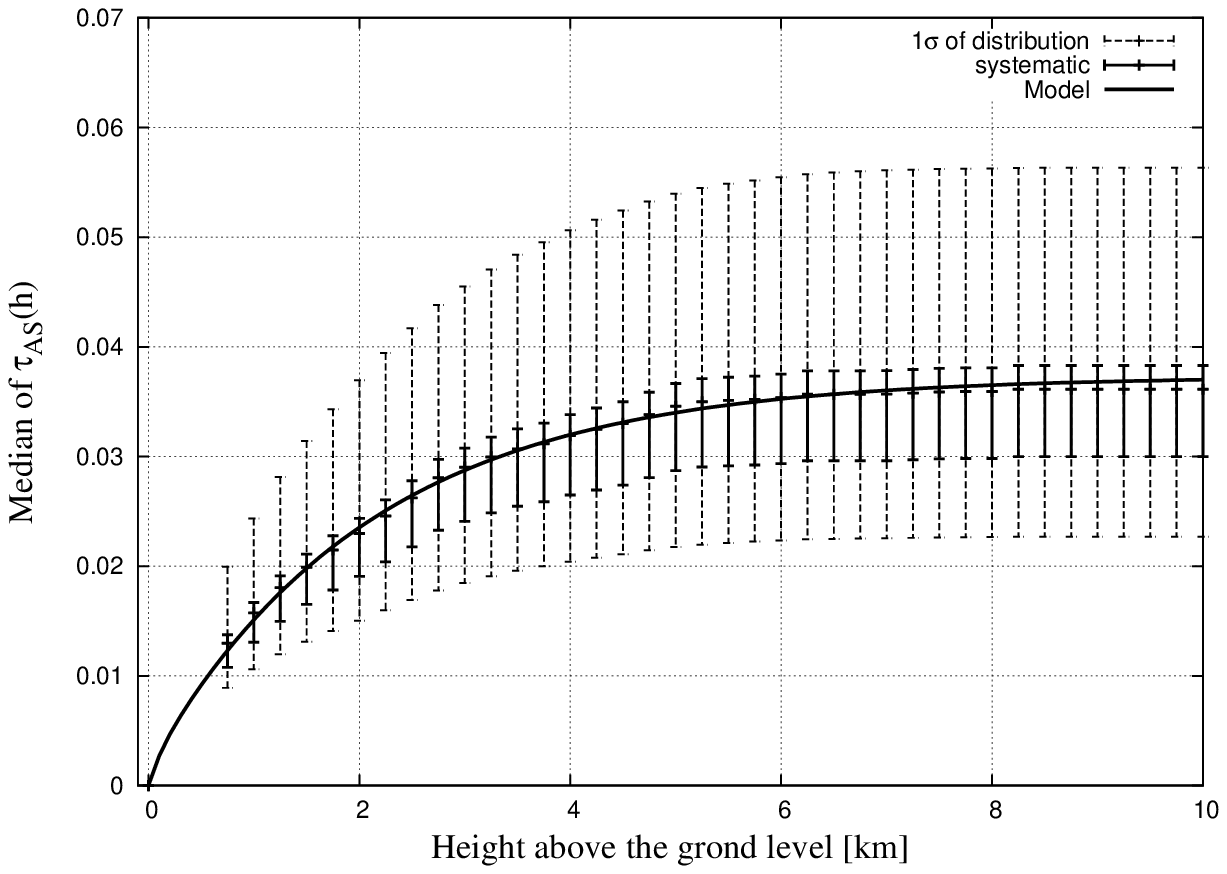}
\caption{Median of $\tau_{\rm AS}(h)$.}
\label{vaod}
\end{center}
\end{minipage}
\begin{minipage}{0.5\textwidth}
\begin{center}
\includegraphics[scale=0.5]{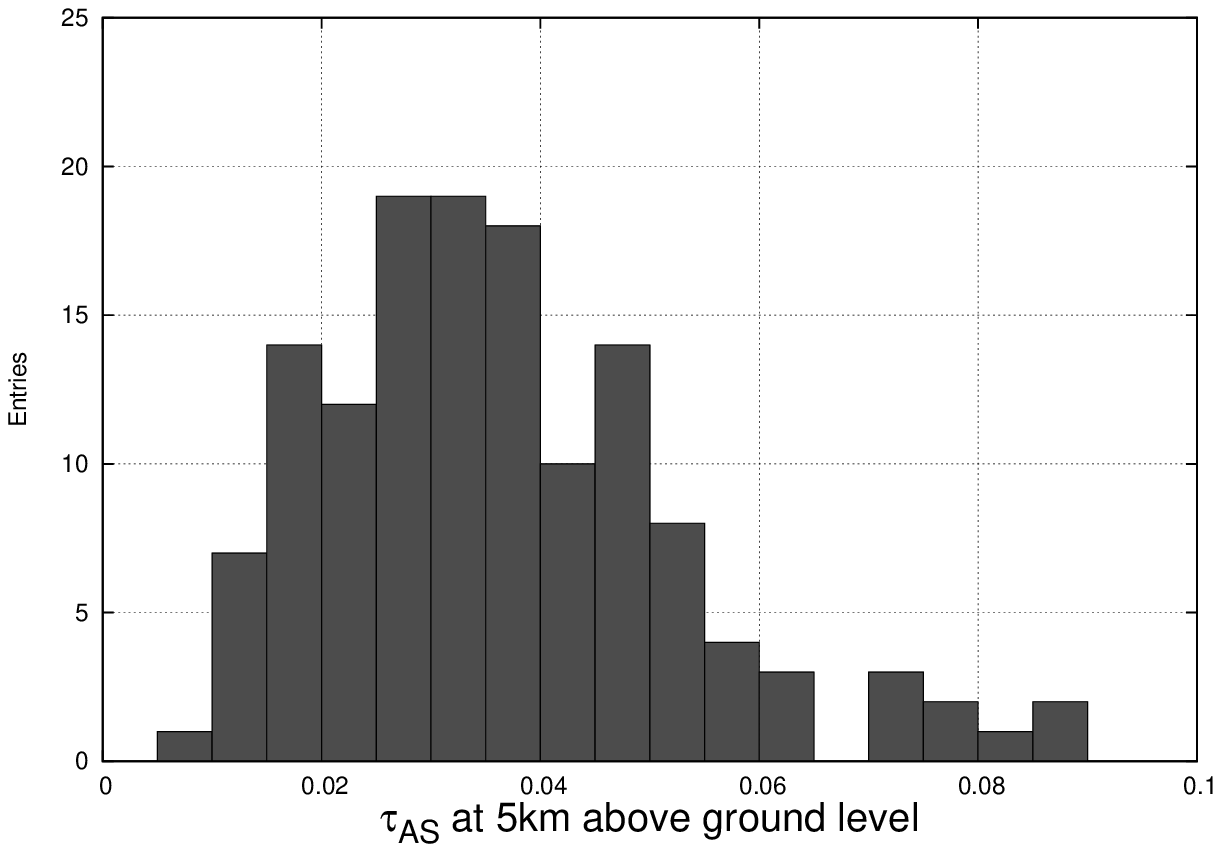}
\caption{Distribution of $\tau_{\rm AS}$ at 5.0km.}
\label{vaod_hist}
\end{center}
\end{minipage}
\end{tabular}
\end{figure}

In this section, we evaluate a transmittance of the atmosphere in terms of the Vertical Aerosols Optical Depth (VAOD).
The VAOD at the height of $h$, $\tau_{\rm AS}(h)$, is defined by the integration of $\alpha_{\rm AS}$ from 0 to $h$ as follows,
\begin{eqnarray}
\tau_{\rm AS}(h) &\equiv& \int_0^h \alpha_{\rm AS}(h') \, {\rm d}h'.
\label{eq:VAOD_equation}
\end{eqnarray}
The attenuation factor for photons which vertically propagate in the atmosphere until height $h$ is given by $\exp \left\{ - \tau_{\rm AS}(h) \right\}.$
The $\tau_{\rm AS}(h)$ was calculated as a function of height using $\alpha_{\rm AS}(x)$ night by night.

In the integration of Eq. ($\ref{eq:VAOD_equation}$), the values of $\alpha_{\rm AS}(h)$ between $h=0$ and $h_{min}$ are assumed to be the average of two curves; 

1. Exponential curve connecting $\alpha_{\rm AS}(0)$ and $\alpha_{\rm AS}(h_{min})$\newline
2. Extrapolation of the exponential curve fitted with the $\alpha_{\rm AS}(h)$ between $h_{min}$ and $h_{min} + 1 \rm km$\newline
where $h_{min}$ is defined as the minimum height for which $\alpha_{\rm AS}$ can be determined by vertical shots.
These two curves are shown by two broken lines in Fig.$\ref{ext_sample}$.

The median of $\tau_{\rm AS}(h)$ is shown in Fig.$\ref{vaod}$.
The distribution of $\tau_{\rm AS}$ at 5km above the ground is shown in Fig.$\ref{vaod_hist}$.
The median of $\tau_{\rm AS}$ at 5km and its range (1$\sigma$) of the distribution are $0.035^{+0.019}_{-0.013}$.
The systematic error of the VAOD was evaluated by taking these two curves as the upper and lower limits, and is estimated to be $\pm6\%$.
Additional systematic errors are $^{+0}_{-16}\%$ from the linearity correction of the PMT response and $^{+2}_{-0}\%$ by the assumption of $\kappa=1.0$ in the Klett's integration method.
The total systematic error of VAOD is $^{+6}_{-17}\%$ added in quadrature.

\begin{figure}[tb]
\begin{center}
\subfigure[Median of $\tau_{\rm AS}(h)$ for different seasons]{
    \includegraphics[width=.5\columnwidth]{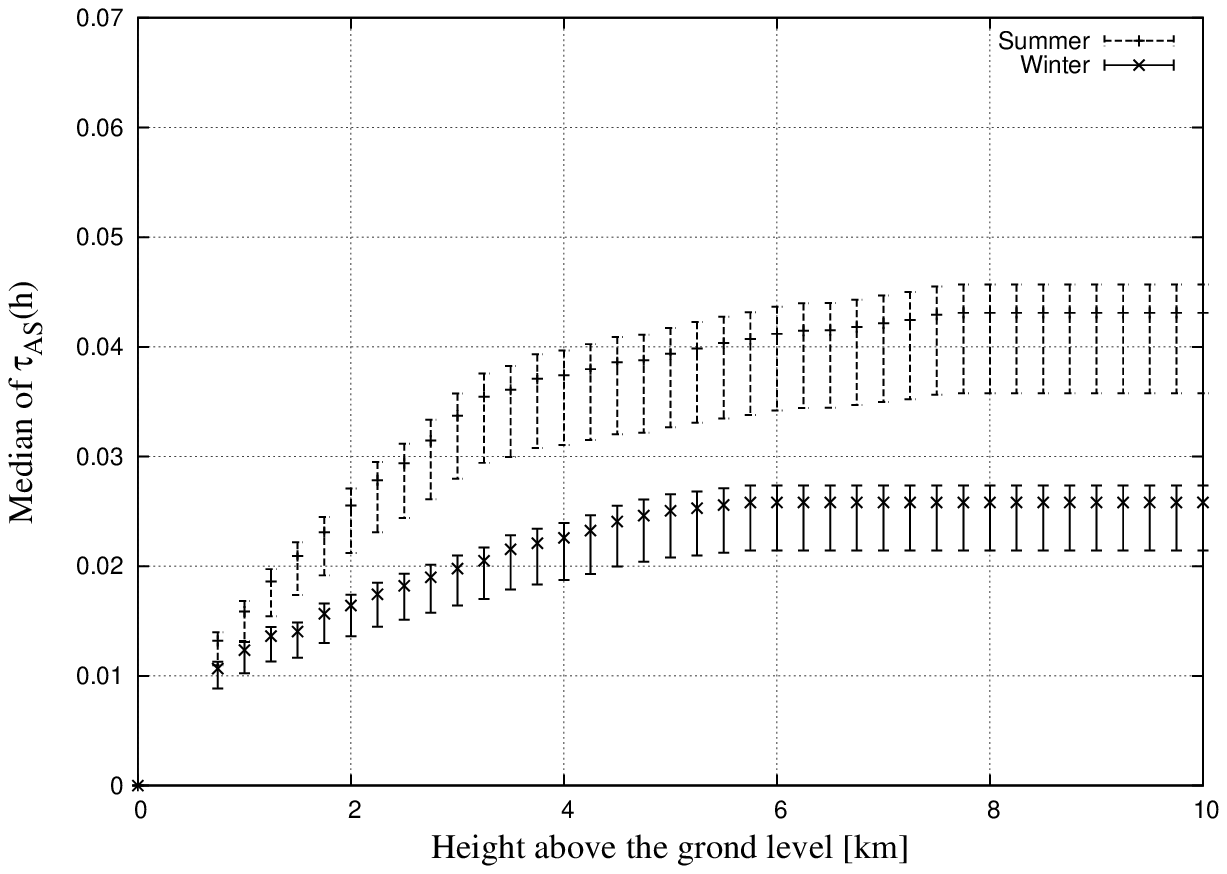}
}~
\subfigure[Distribution of $\tau_{\rm AS}$ at 5\rm km above ground level for summer (above) and winter (below)]{
    \includegraphics[width=.5\columnwidth]{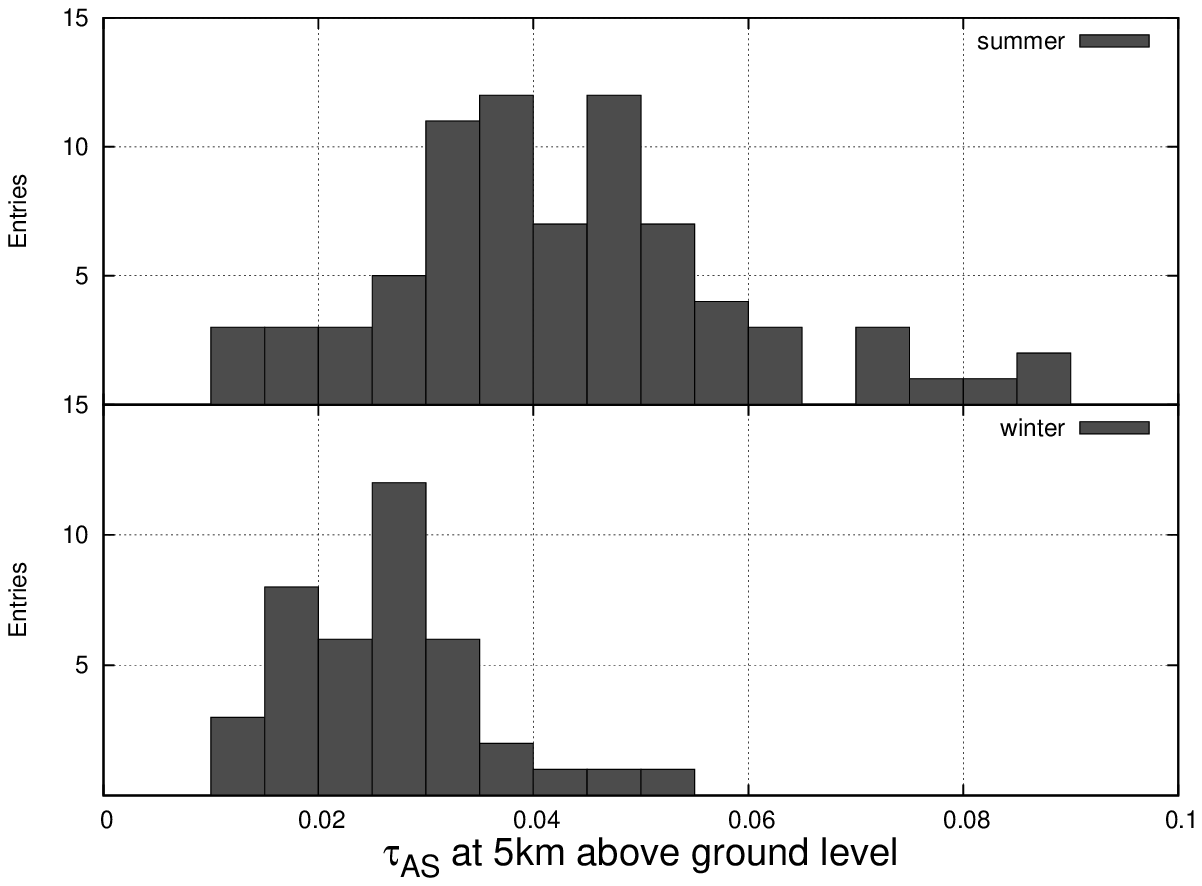}
}
\caption{Aerosol scattering status for different seasons.}
\label{according_to_season}
\end{center}
\end{figure}

The seasonal variation of $\tau_{\rm AS}$ distribution is shown in Fig.$\ref{according_to_season}$.
The data of winter and summer include the data for five months that centered January and July, respectively.
The mean values of $\tau_{\rm AS}$ and their distribution range (1 $\sigma$) are $0.039^{+0.020}_{-0.012}$ in summer and $0.025^{+0.010}_{-0.007}$ in winter.
The effect of the aerosol component in summer is 1.5 times greater than that in winter.

\section{Model of transparency distribution}
\label{sec:Model}
\begin{figure}[tb]
\begin{center}
\subfigure[linear scale]{
    \includegraphics[width=.5\columnwidth]{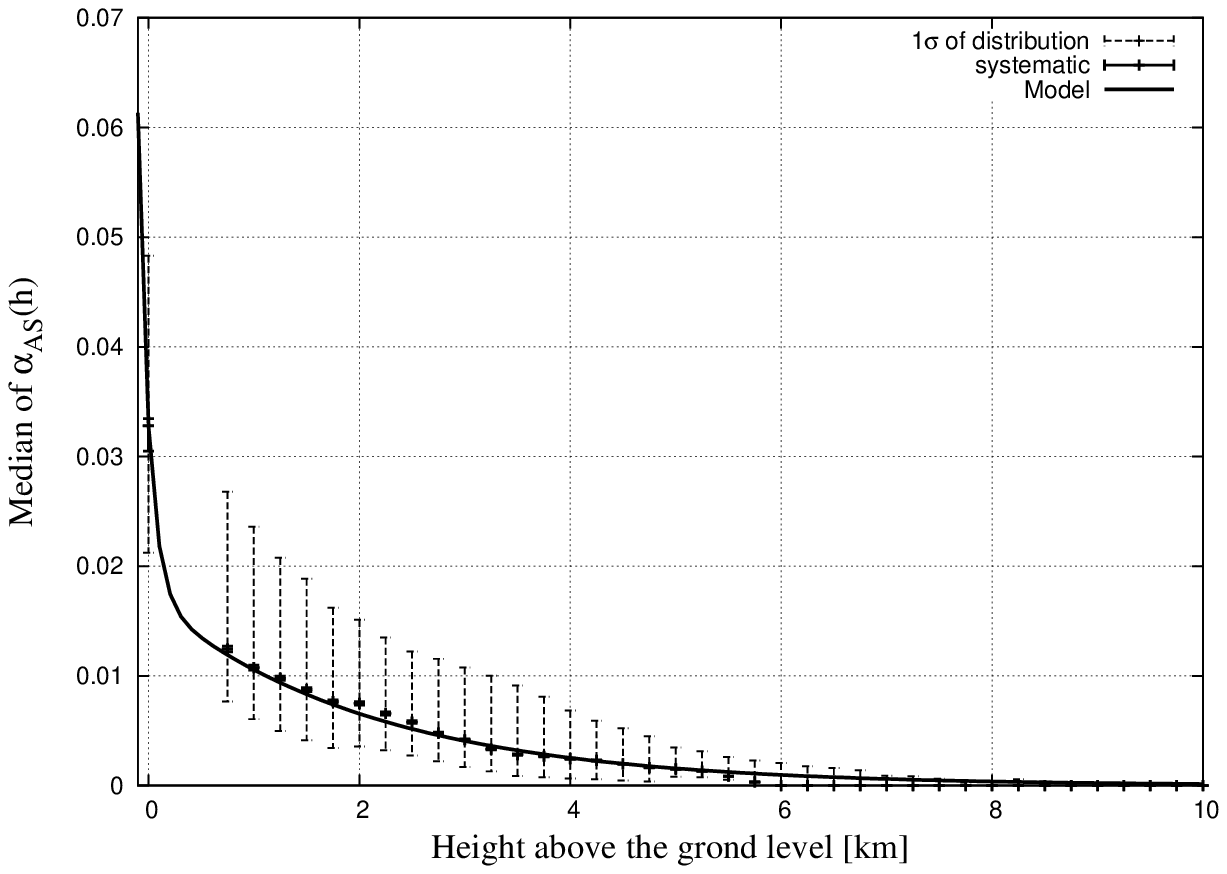}
}~
\subfigure[log scale]{
    \includegraphics[width=.5\columnwidth]{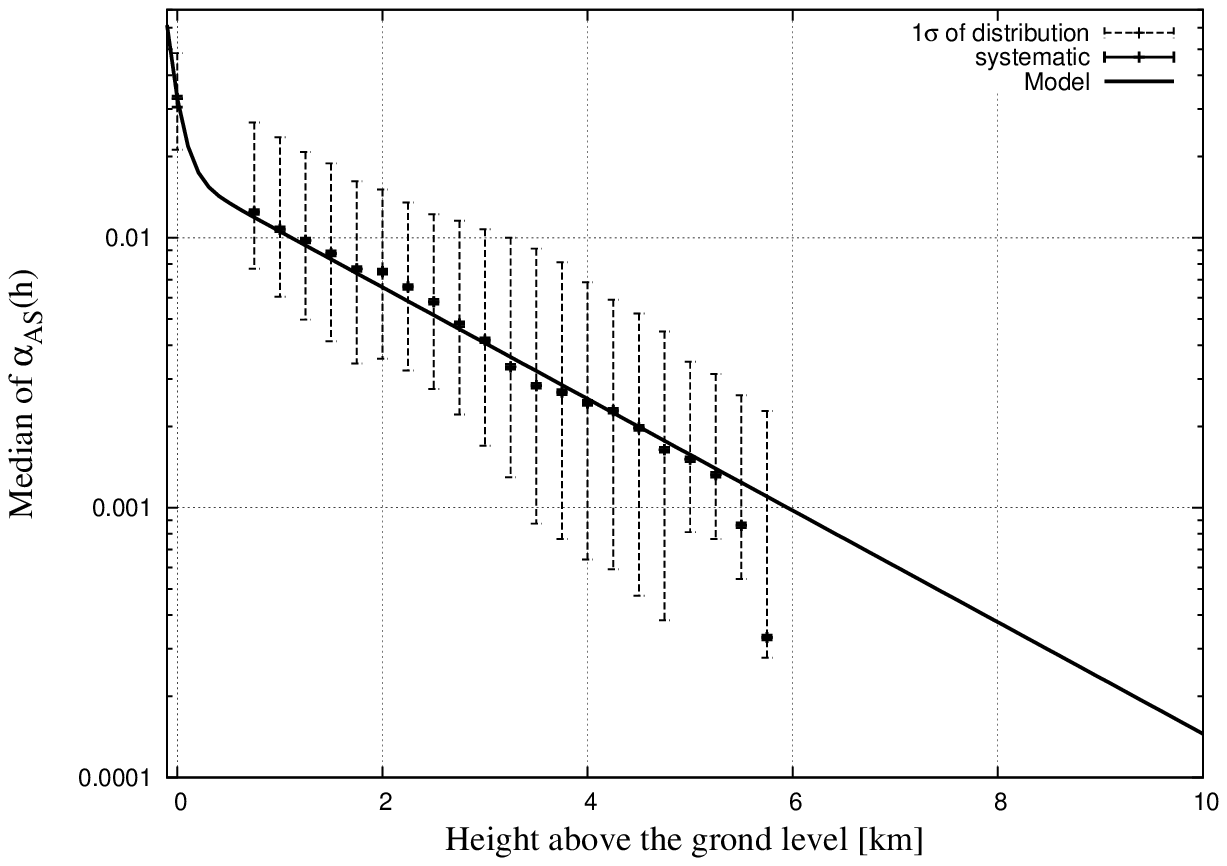}
}
\caption{
    Distribution of $\alpha_{\rm AS}$ as a function of the height from the ground level.
    At each height, crosses show a median of $\alpha_{\rm AS}$, broken error bars indicate the range (68\%) of its distribution.
}
\label{ext}
\end{center}
\end{figure}

In this section, a model of the altitudinal distribution of the atmospheric transparency based on the $\alpha_{\rm AS}(h)$ measurement is presented.
Figure $\ref{ext}$ shows the distribution of $\alpha_{\rm AS}$ averaged over 137 LIDAR operations as a function of the height $h$ above the ground level.
The median of $\alpha_{\rm AS}$ shown in Fig.$\ref{ext}$ can be well expressed by two-component exponential function;
\begin{eqnarray}
\alpha_{\rm AS}(h)=0.017\exp(-h/2.1)+0.016\exp(-h/0.1)
\label{eq:AS_model}
\end{eqnarray}
where $h$ is the height from the ground level of LIDAR observation ($1400 \rm m$ a.s.l.) in the unit of $\rm km$.
The first term of the function ($\ref{eq:AS_model}$) is responsible for the distribution of the aerosol at high altitude ($h\ge0.75 \rm km$) whereas the lower altitude distribution is expressed by the super position of two terms.
Especially at the ground level ($h\sim 0$), two terms have nearly equal contributions.
The functional from of $\alpha_{\rm AS}(h)$ is plotted in Fig.$\ref{ext}$ as the solid line.
The integral of $\alpha_{\rm AS}(h)$ is overlaid as a solid line in Fig.$\ref{vaod}$ and it reproduces the median value of measured $\tau_{\rm AS}(h)$ well.

\section{Conclusion}
\label{sec:Conclusion}
The LIDAR atmospheric monitoring system of TA began operation in September 2007 and continues until today.
Analysing the data from the LIDAR, we have found that the extinction coefficient by aerosol at the ground level is $\alpha_{\rm AS}(0 \rm km)= 0.033^{+0.016}_{-0.012} \rm km^{-1}$ corresponding to $30^{+16}_{-10} \rm km$ in the attenuation length where the error ($\pm$) corresponds to the range of distribution ($1 \sigma$).
The maximum altitude at which the contribution of the aerosol scattering was observed, excluding the cloud, is $5 \rm km$.
We obtained the VAOD at each altitude $h [\rm km]$, integrating $\alpha_{\rm AS}(h)$ from 0 to $h$.
The median of the VAOD at 5 km above the ground is $\tau_{\rm AS}(5 \rm km)= 0.035^{+0.019}_{-0.013}$ with a systematic error of $^{+6}_{-17}\%$.
The effect of the aerosol in summer is found to be $1.5$ times greater than that in winter at the observation site.
The average altitudinal distribution of $\alpha_{\rm AS}$ above the ground level ($1400 \rm m$ a.s.l.) is well reproduce by the super position of two exponential functions;
\[\alpha_{\rm AS}(h)=0.017\exp(-h/2.1)+0.016\exp(-h/0.1).\]

\section*{Acknowledgments}
The Telescope Array experiment is supported
by the Ministry of Education, Culture, Sports, Science and Technology-Japan
through Kakenhi grants on priority area ($\sharp$431) ``Highest Energy Cosmic Rays'';
by the U.S. National Science Foundation awards PHY-0307098, PHY-0601915, PHY-0703893, PHY-0758342, and PHY-0848320 (Utah) and PHY-0649681 (Rutgers);
by the Korea Research Foundation (KRF-2007-341-C00020); by the Korean Science and Engineering Foundation (KOSEF, R01-2007-000-21088-0);
A Korean WCU grant (R32-2008-000-10130-0) from MEST and the National Research Foundation of Korea(NRF).
The experimental site became available through the cooperation of the Utah School and Institutional Trust Lands Administration (SITLA), U.S.~Bureau of Land Management and the U.S.~Air Force.
We also wish to thank the people and the officials of Millard County, Utah, for their steadfast and warm supports.

\end{document}